\newcommand{\showfigs}[1]{#1} 
\newcommand{\myslug}[1]{} 
\newcommand{\blstretch}{1} 

\documentstyle[11pt,agums,epsf,graphics]{article} \renewcommand{\blstretch}{2} \renewcommand{\myslug}[1]{#1} \renewcommand{\showfigs}[1]{}

\lefthead{O'BRIEN ET AL.}
\righthead{STATISTICAL ASYNCHRONOUS REGRESSION}
\received{2000}
\revised{2000}
\accepted{2000}
\paperid{} 
\cpright{AGU}{1998}
\ccc{0148-0227/97/97GL-00000\$05.00}
\authoraddr{T.P.\ O'Brien, D.\ Sornette, R. L. McPherron,
Institute of Geophysics and Planetary Physics, UCLA, 405 Hilgard, Los
Angeles, CA 90095-1567,
tpoiii@igpp.ucla.edu, sornette@moho.ess.ucla.edu, 
\protect{\linebreak} rmcpherr@igpp.ucla.edu}
\newcommand{\Nx}{15} 
\newcommand{\Ny}{25}

\begin{document}

\title{Statistical Asynchronous Regression: Determining the Relationship \\
Between two Quantities that are not Measured Simultaneously}

\author{T.P.\ O'Brien$^{\rm 1}$, D.\ Sornette$^{\rm 1,2,3}$ and R.L.\
McPherron$^{\rm 1,2}$}

\affil{1. Institute of Geophysics and Planetary Physics, UCLA, Los Angeles,
CA 90095}
\affil{2. Department of Earth and Space Sciences, UCLA, Los Angeles, CA, 90095}
\affil{3. Laboratoire de Physique de la Mati\`{e}re Condens\'{e}e, CNRS and
Universit\'{e} de Nice-Sophia Antipolis, 06108 Nice Cedex 2, France}

\myslug{
\vfill
\noindent
IGPP Publication No. 5471

\vfill
\noindent
\today
}

\begin{abstract}
We introduce the Statistical Asynchronous Regression (SAR) method: a technique
for determining a relationship between two time varying quantities without
simultaneous measurements of both quantities. We require that there is a time 
invariant, monotonic function Y = u(X) relating the two quantities, Y and X.
In order to determine u(X), we only need to know the statistical distributions
of X and Y. We show that u(X) is the change of variables that converts the 
distribution of X into the distribution of Y, while conserving probability.
We describe an algorithm for
implementing this method and apply it to several example distributions. We also
demonstrate how the method can separate spatial and temporal variations from a
time series of energetic electron flux measurements made by a spacecraft in
geosynchronous orbit. We expect this method will be useful to the general
problem
of spacecraft instrument calibration. We also suggest some applications of 
the SAR method outside of space physics.
\end{abstract}


\begin{article}
\renewcommand{\baselinestretch}{\blstretch}
\section{Introduction}

We developed the Statistical Asynchronous Regression (SAR) 
technique described in this paper 
as part of a study of relativistic electron conditions 
at geosynchronous orbit. This part of the Earth's radiation 
belts can evolve on a timescale of hours or even minutes. Unfortunately,
while individual satellites may make measurements every few seconds,
it is difficult to separate the temporal changes from consequences of
orbital motion. The easiest way to do this would be to have continuous
measurements at a fixed location, or local time, such as local noon.
Instead, we have continuous measurements on board moving spacecraft.
We can remove the orbital effects if we can map our continuous
measurements to local noon at geosynchronous orbit. 

Relativistic electrons in the vicinity
of geosynchronous orbit drift around the earth every 5-15 minutes under
the influence of the local magnetic field. As it happens, these electrons
do not follow circular paths like satellite orbits, but rather elliptical
paths that depend on the details of the local magnetic field geometry. However,
because electron density is a relatively smooth function of 
altitude near geosynchronous orbit, measurements at different local times
are strongly correlated. This correlation is stronger still if we average our
data over several drift periods. The strong correlation suggests that we
can map our continuous measurements to local noon, if we can determine
the right mapping function.

Sometimes it is possible to determine empirical 
mappings between measurements at different local times by regression
of simultaneous measurements. For example, it is possible to relate
measurements made by the GOES~8 spacecraft at local dawn (0600) 
to GOES~9 measurements at local
10 AM (1000), because whenever GOES~8 is at local dawn, GOES~9 is at
local 10 AM. However, it is never the case that GOES~8 is at 
local dawn when GOES~9 is at local noon. Therefore, we need some method 
for mapping measurements from anywhere to local noon (or some other local 
time of interest). Until recently, there have been three strategies for
resolving this difficulty: interpolate between multiple calibrated 
spacecraft \cite{Reeves1998}, use the equation of motion of electrons 
in model electromagnetic fields to follow particles around 
geosynchronous orbit \cite{friedel99},
or use some kind of empirical description of the orbital 
variations \cite{CRRESELE,ae8min}. The first approach degrades 
substantially when only a few
spacecraft are available, and fails when only one spacecraft is available.
The second approach suffers from the substantial imperfections in our
magnetic field models near geosynchronous \cite{Selesnick2000}. 
The third approach has been
applied with encouraging success by {\it Moorer}\nocite{Moorer1999} [1999], 
who uses
whatever measurements are available to adjust the CRRESELE empirical 
radiation belt model for best agreement. The SAR technique provides us
with a more robust approach that can be applied in cases when there is
no pre-existing empirical model like CRRESELE.
The SAR technique calibrates not only between spacecraft and 
instruments but also between different locations (local times) around 
geosynchronous orbit. One can easily imagine the
SAR technique as calibrating measurements made by GOES~8 at local dawn
to measurements made by GOES~9 at local noon--even though these two 
spacecraft have never been at these locations simultaneously. Additionally,
the SAR technique is non-parametric because it does not require us to assume
a functional form for the mapping between local times.

When we have described the SAR technique to our colleagues, 
many have found it novel and challenging to understand, and some have
stated that it might be useful in their own work on other problems. 
For our own purposes, since we have used this technique as the basis 
of a statistical study of the energetic electrons near geosynchronous 
orbit, we present this technique to familiarize our audience 
with the technique and to demonstrate its robustness. As we believe 
the SAR technique has applications beyond the electron radiation belts, 
we have chosen to dedicate this paper entirely to the technique itself, 
reserving the radiation belt study to a later publication. 

In essence, our method provides a means of performing a regression
of one time varying quantity against another without requiring simultaneous 
knowledge of both. We call this the Statistical Asynchronous Regression (SAR) method,
because it allows us to regress $Y(t)$ against $X(t)$
using only the two statistical distributions $F(x)$ and $G(y)$. The SAR 
method determines the
function $Y = u(X)$ by matching the quantiles (or percentiles) $x$ and $y$ of 
the distributions of X and Y for each probability level. 
A primitive variant of this
technique was developed to standardize the calculation of $K$ indices at
different magnetic observatories \cite[and references therein]{mayaud80}.
We also note that a transformation similar to the SAR method has been 
introduced to map non-Gaussian random variables onto Gaussian ones, 
with application to the construction of multivariate distribution 
functions in high-energy particle physics experiments \cite{karlen}, 
in the theory of portfolio in Finance \cite{portfolio}, and
earlier in the treatment of bivariate gamma distributions \cite{Moran}.

In statistics, one method of graphical hypothesis testing is the Q-Q 
(quantile-quantile) plot \cite{Wilk68}, which is essentially a
graphical depiction of $u(X)$ based on the same principle as the SAR method. 
A linear $u(X)$ indicates that
the two variables differ only by a scaling and an offset but are
otherwise identically distributed. 
However, in spite of the variety of graphical techniques related to 
the SAR method, none makes use of the plotted $u(X)$, aside from determining 
whether it is linear \cite{Fisher83}. Since we are specifically 
interested in potentially nonlinear $u(X)$, we have developed the SAR
method as an extension to the Q-Q plot. 

Under various names,
such as \emph{anchoring} or the \emph{equipercentile} method,
psychological and educational testing use the same principle 
as the SAR technique to normalize a new test to 
a standard score distribution \cite{Allen1979}. However,
$u(X)$ is not explicitly calculated, and the information
it contains is typically discarded.

Additionally, the Spearman rank
order correlation coefficient touches on the same notion as the SAR 
method \cite{Press92}.
It calculates a linear correlation coefficient between the 
sorted rank orders of two quantities rather than the quantities 
themselves; this coefficient measures the quality of the optimal
nonlinear mapping between two simultaneously measured quantities.
Since we are concerned with comparing quantities not measured 
simultaneously, we will not make use of the Spearman coefficient.

In the remainder of this paper, we will provide a description and 
some limited analysis of the SAR method. First, we 
will describe the technique by parable, using a graphical illustration. 
Next we will provide the formal derivation of the technique. We will provide 
several examples and a simple recipe for the implementation of the SAR 
technique. Then we will address the problems of finite sample size and 
noisy measurements. Finally we will show how we use the SAR method to
map geosynchronous energetic electron flux from one local time to another.

\section{A Simple Example}
We begin our explanation of the SAR technique by taking a step back from
space physics to a simpler analogous problem. Suppose
we have two meteorologists making measurements every other day. One has been
measuring his favorite meteorological quantity $X$, and the other has been
measuring $Y$.
Unfortunately, owing to an error in scheduling, the two meteorologists have not
been making their measurements on the same days. It is therefore impossible for
them to plot $Y$ against $X$ and perform a regression. We will show how it is
nonetheless possible for them to recover the empirical function $Y = u(X)$. The
powerful statistical tool that will make this possible is the fundamental
principle that probability is conserved under a change of variables. We will
leave the mathematical presentation of this principle to later sections.

In \callout{Figure~\ref{PedanticPDF}}, we have plotted the probability density
functions (PDFs) $f(x)$ and $g(y)$ along the $x$- and $y$-axes respectively.
For clarity, we have plotted $f(x)$ upside down and $g(y)$ rotated
counterclockwise.
Each density function represents the distribution of observations made by one
of the
scientists. In this example, $X$ is distributed uniformly between 1 and 2, and
$Y$ is distributed as $1/y$ between $e$ and $e^{2}$. We have also plotted
the relational
function $Y = u(X) = e^{X}$ that provides the change of variables. The shaded
area within $f(x)$ is the probability that a single measurement of $X$ falls
between $x_{1}$ and $x_{2}$. Similarly, the shaded area within $g(y)$ is the
probability that a single measurement of $Y$ falls between $y_{1}=u(x_1)$
and $y_{2}=u(x_2)$.
The conservation of probability is illustrated graphically by the fact that the
two shaded regions are equal in area. With any two of these three curves, it is
possible to determine the third. Generally, it
has been of greater interest to reconstruct $g(y)$ knowing $f(x)$ and
$u(X)$. We,
however, are interested in reconstructing $u(X)$ knowing only $f(x)$ and
$g(y)$. The fundamental assumption is that of
stationarity: the unknown relationship $Y = u(X)$ is the same at all times;
this condition must be met for a statistical approach to be possible.

\begin{figure}
\showfigs{
\newpage
\centerline{\epsffile{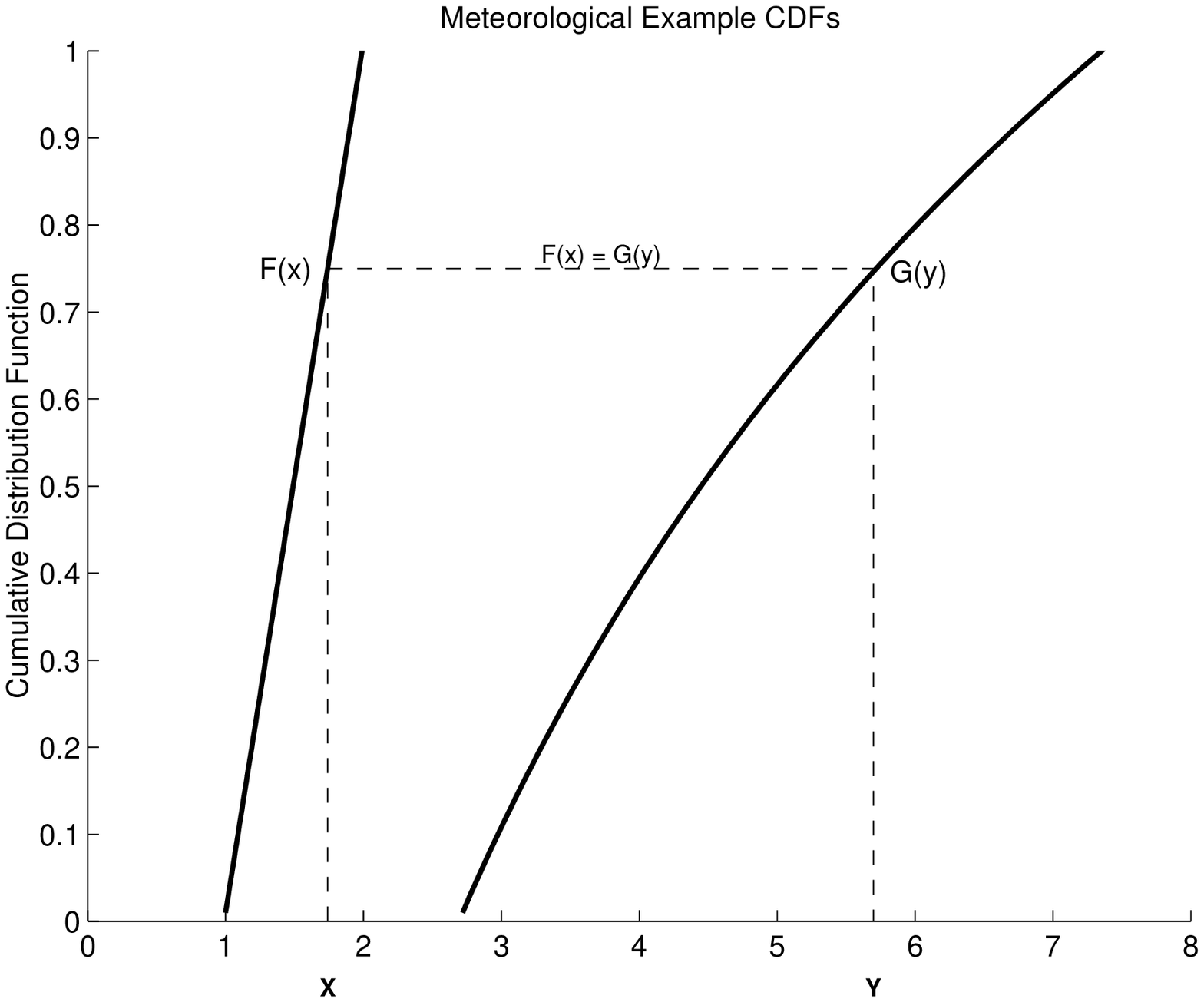}}
}

\caption{Cumulative distribution functions are plotted for $X$ and $Y$ on the
horizontal axis. Following the dashed line, one can easily determine what
value of $Y$ corresponds to a given $X$.}
\label{PedanticCDF}
\end{figure}

One can reconstruct $Y = u(X)$ for each $X$ simply by finding the value $Y$
such
that the area inside $g(y)$ from $-\infty$ to $Y$ is equal to the area inside
$f(x)$ from $-\infty$ to $X$. In \callout{Figure~\ref{PedanticCDF}} we
demonstrate this cumulative way of looking at the problem. Instead of plotting
the density functions $f(x)$ and $g(y)$, we have plotted the cumulative
distribution functions (CDFs) $F(x)$ and $G(y)$. The CDFs are the integrals
from
$-\infty$ to $x$ of $f(x)$ and $-\infty$ to $y$ of $g(y)$, and they
correspond to
the areas inside $f(x)$ and $g(y)$ mentioned above. To find the $Y$ that
corresponds to a given $X$ in Figure~\ref{PedanticCDF}, one reads from the $X$
value on the abscissa up to $F(x)$ then horizontally over to the same
value of
$G(y)$, and back down to the abscissa to find the corresponding $Y$.
Compared to Figure~\ref{PedanticPDF}, this visualization makes it easier to
find
$Y$ for a given $X$, but does not provide an obvious representation of $u(X)$.
While emphasizing different features of the method, these two graphical
representations of the method give identical results. In the following
sections,
we will provide the formal mathematical treatment of the graphical operations.

\section{Formalism}

Some of our readers will no doubt be a bit rusty in the manipulation of
probabilities. Therefore, we have included a thorough treatment of the change
of variables theorem in an appendix. Here, we begin with the differential form
of the change of variables:
\begin{eqnarray}
f(x)dx &=& g(u(x))|u'(x)|dx \nonumber \\* 
	 &=& g(y) \left| \frac{dy}{dx} \right| dx = g(y)|dy|. \label{ChangeOfVars}
\end{eqnarray}

In order to use this equation, we must determine the sign of $u'(x)$. For
distributions with only one tail, we can do this rather easily by examining the
rare values of $X$ and $Y$. When the rare values of $X$ and $Y$ fall at the
same
end of the real number line, $u'(x)$ is positive. When they fall at
opposite ends, $u'(x)$ is negative. Physical insight is also a
useful tool
in determining the sign of $u'(x)$. If we expect larger (or more positive)
values
of $X$ to correspond to larger values of $Y$, then $u'(x)$ is
positive. If
we expect larger values of $X$ to correspond to smaller (or more negative)
values
of $Y$, then $u'(x)$ is negative. 

For $u'(x) > 0$, we can integrate (\ref{ChangeOfVars}),
\begin{equation}
\int_{-\infty}^{x} f(x')dx' = \int_{-\infty}^{y} g(y')dy'. \label{UposIntegral}
\end{equation}
\noindent
This equation implicitly defines $y = u(x)$ as the function that provides
the matching integration bounds. We recognize these integrals as the CDFs
of $X$ and $Y$, so we can rewrite (\ref{UposIntegral}) as
\begin{equation}
F(x) = G(y) ~~{\rm for~} u'(x) > 0.
\end{equation}
\noindent
We can invert $G(y)$ to arrive at an explicit equation for $u(x)$,
\begin{equation}
y = G^{-1} \left( F(x) \right) = u(x). \label{UforUpos}
\end{equation}
\noindent
This equation represents the mathematical counterpart to the graphical
operation
described in Figure~\ref{PedanticCDF}, where one moves up from $X$ to $F(x)$,
then across to $G(y)$, then back down to the corresponding $Y$.

For $u'(x) < 0$, we can integrate (\ref{ChangeOfVars}),
\begin{equation}
\int_{-\infty}^{x} f(x')dx' = \int_{y}^{+\infty} g(y')dy'. \label{UnegIntegral}
\end{equation}
\noindent
Converting this equation to CDFs, we have
\begin{equation}
F(x) = 1-G(y) ~~{\rm for~} u'(x) < 0.
\end{equation}
\noindent
solving for $u(x)$, we arrive at
\begin{equation}
y = G^{-1} \left( 1-F(x) \right) = u(x). \label{UforUneg}
\end{equation}
\noindent
Combining (\ref{UforUpos}) and (\ref{UforUneg}) we arrive at
\begin{equation}
u(x) = \left\{ \begin{array}{ll}
		G^{-1}(F(x)) & {\rm for}~~u'(x) > 0, \\*
		G^{-1}(1-F(x)) & {\rm for}~~u'(x) < 0.
		  \end{array}
	 \right.
\label{Ufinal}
\end{equation}

It is clear, then, that all we need to determine $u(x)$ is knowledge of the
sign of $u'(x)$ and either $F(x)$ and $G(y)$ or $f(x)$ and $g(y)$. We summarize
the desirable properties of $u(x)$ as follows:
\noindent
\begin{itemize}
\item it can be arbitrarily nonlinear;
\item its determination is not parametric;
\item it maps the entire distribution and all of the moments of $X$ onto
those of $Y$;
\item it can be determined without simultaneous measurements of $X$ and $Y$.
\end{itemize}

\section{More Examples}
We now turn to some more sophisticated examples of the SAR method. First,
we will
return to our original meteorological example to demonstrate the SAR
procedure on
analytical functions. Then, we will provide a function relating a bimodal
distribution to a Gaussian. Finally, we will demonstrate the method on a
stretched exponential and a Gaussian.

\subsection{Meteorological Example}
In the example of the meteorologists, illustrated in Figures
\ref{PedanticPDF} and \ref{PedanticCDF},
the following analytical functions were used:
\begin{eqnarray}
f(x) &=& \left\{ \begin{array}{ll}
		1 & {\rm for}~~1 \leq x < 2, \\*
		0 & \mbox{otherwise},
		  \end{array}
	 \right.  \label{AnalyticalPDFx} \\
g(y) &=& \left\{ \begin{array}{ll}
		1/y & {\rm for}~~e \leq y < e^{2}, \\*
		0 & \mbox{otherwise} .
		  \end{array}
	 \right.  \label{AnalyticalPDFy}
\end{eqnarray}
\noindent
Using (\ref{xCDF}) and (\ref{yCDF}) together with (\ref{AnalyticalPDFx})
and (\ref{AnalyticalPDFy}), we have
\begin{eqnarray}
F(x) &=& \left\{ \begin{array}{ll}
		0 & {\rm for}~~x < 1, \\*
		x-1 & {\rm for}~~1 \leq x < 2, \\*
		1 & {\rm for}~~x \geq 2,
		  \end{array}
	 \right.  \label{AnalyticalCDFx} \\
G(y) &=& \left\{ \begin{array}{ll}
		0 & {\rm for}~~y < e, \\*
		\log y - 1 & {\rm for}~~e \leq y < e^{2}, \\*
		1 & {\rm for}~~y \geq e^{2}.
		  \end{array}
	 \right. \label{AnalyticalCDFy}
\end{eqnarray}
Inserting (\ref{AnalyticalCDFx}) and (\ref{AnalyticalCDFy}) into
(\ref{UforUpos}), we see that
\begin{equation}
u(x) = G^{-1}(F(x)) = e^{1+F(x)} = e^{x}.
\end{equation}
\noindent
Adding in the proper bounds, we have
\begin{equation}
u(x) = e^{x} ~{\rm for}~~ 1 \leq x < 2.
\end{equation}

\subsection{Bimodal Example}

In our next example, we will show how the SAR method easily handles bimodal
distributions. We have chosen $X$ to be bimodal and $Y$ to be unimodal. The
PDFs are
\begin{eqnarray}
\lefteqn{f(x) =} \nonumber \\*
& & \frac{1}{2\sqrt{2\pi}}\left(e^{-\case{1}{2}(x-3)^{2}}+e^{-\case{1}{2}(x-8)^{2}}
\right),
\end{eqnarray}
\begin{equation}
g(y) = \frac{
1}{3\sqrt{2\pi}}e^{-\frac{1}{2}\left(\frac{y-10}{3}\right)^{2}}.
\end{equation}
\noindent
While there is no closed form for $u(X)$, a graphical display can show its
qualitative features. \callout{Figure~\ref{BimodalPDF}} shows how the bimodal
$f(x)$ maps to $g(y)$. The highly nonlinear mapping $u(X)$ has a flat spot
(with small but
still  positive slope)
corresponding to the local minimum in $f(x)$, since $u'(x) = f(x)/g(u(x))$.
In Figure~\ref{BimodalPDF}, we see
how a large range of $X$ values near $X=5$ maps to a very narrow range of $Y$
values near $Y=10$. More generally, the terraced shape of $u(X)$ can be seen
to generate bimodal or multimodal distributions from unimodal ones.

\subsection{Stretched Exponential Example}

For our final example, we will treat an unusual distribution and an unusual
mapping. We consider the case of a stretched exponential mapped to a
Gaussian. In
this case, $X$ and $Y$ are distributed as
\begin{eqnarray}
\lefteqn{f(x) =} \nonumber \\*
& & \frac{c}{\sqrt{\pi}x_{0}}\left(\frac{x}{x_{0}}\right)^{\case{c}{2}-1}
e^{-\left(\frac{x}{x_{0}}\right)^{c}}~~{\rm for}~x > 0,  \label{stretchPDF}
\end{eqnarray}
\begin{equation}
g(y) = \sqrt{\frac{2}{\pi\sigma^{2}}} e^{-\frac{(y-\mu)^{2}}{2\sigma^{2}}}
~~{\rm for}~y > \mu,
\end{equation}
\noindent
where $c$, $\sigma$, and $x_{0}$ are positive real values. Using
(\ref{ChangeOfVars}) and assuming $u'(x)>0$, we can write a differential
equation
for $u(x)$,
\begin{eqnarray}
\frac{c}{\sqrt{\pi}x_{0}}\left(\frac{x}{x_{0}}\right)^{\case{c}{2}-1}
e^{-\left(\frac{x}{x_{0}}\right)^{c}} = \nonumber \\*
\sqrt{\frac{2}{\pi\sigma^{2}}}e^{-\frac{(u(x)-\mu)^{2}}{2\sigma^{2}}}u'(x).
\label{Upowerlaw}
\end{eqnarray}
\noindent
By our design of (\ref{stretchPDF}),
 $u(x)$ will cause the two exponentials to drop out of the equation,
satisfying the system
\begin{eqnarray}
-\left(\frac{x}{x_{0}}\right)^{c} &=& -\frac{(u(x)-\mu)^{2}}{2\sigma^{2}},
\label{ExponentialsU} \\*
\frac{c}{\sqrt{\pi}x_{0}}\left(\frac{x}{x_{0}}\right)^{\case{c}{2}-1} &=&
\sqrt{\frac{2}{\pi\sigma^{2}}}u'(x). \label{ExponentialsUprime}
\end{eqnarray}
\noindent
Solving (\ref{ExponentialsU}) for $u(x)$ we have
\begin{equation}
u(x) = \sqrt{2}\sigma\left(\frac{x}{x_{0}}\right)^{\case{c}{2}} + \mu,
\end{equation}
\noindent
which is, in fact, the solution to (\ref{ExponentialsUprime}) and thus of
(\ref{Upowerlaw}).
This mapping function is a highly nonlinear
power-law. In \callout{Figure~\ref{StretchPDF}}, we have depicted the
borderline case for $c = 1$, $x_{0}=1$, $\sigma = 1/\sqrt{2}$ and 
$\mu=0$. For $c<1$, this distribution becomes a stretched exponential,
which is a common distribution in real data. While $f(x)$
diverges at $x=0$, the SAR method cleanly recovers the mapping function $u(x) =
2\sqrt{x}$. We are now going to investigate the robustness of the SAR
method on finite and noisy data sets.

\section{The Algorithm and Associated Approximation Problems}

So far, we have considered the analytical representations of $f(x)$ and $g(y)$.
However, in practice, we will only have a finite number of samples of each 
variable. We can use these samples to construct $F(x)$ and $G(y)$ and then
perform either a tabular or an analytical approximation to (\ref{Ufinal}).

First, we sort the $X$ and $Y$ values. These sorted values give us an 
approximation to $F(x)$ and $G(y)$. For example, if $x_{i}$ is the
$i^{\rm th}$ smallest value in $N_{x}$ measurements of $X$, then an estimate of
$F(x)$ is
\begin{equation}
F^{\ast}(x_{i}) = \frac{i}{N_{x}}.
\label{Fest}
\end{equation}
\noindent
Similarly, we estimate $G(y)$ as
\begin{equation}
G^{\ast}(y_{j}) = \frac{j}{N_{y}}.
\label{Gest}
\end{equation}
\noindent
There are more sophisticated methods of estimating these distributions,
such as kernel estimators \cite{Hardle}, if the need arises.

\begin{figure}
\showfigs{
\newpage
\centerline{\epsffile{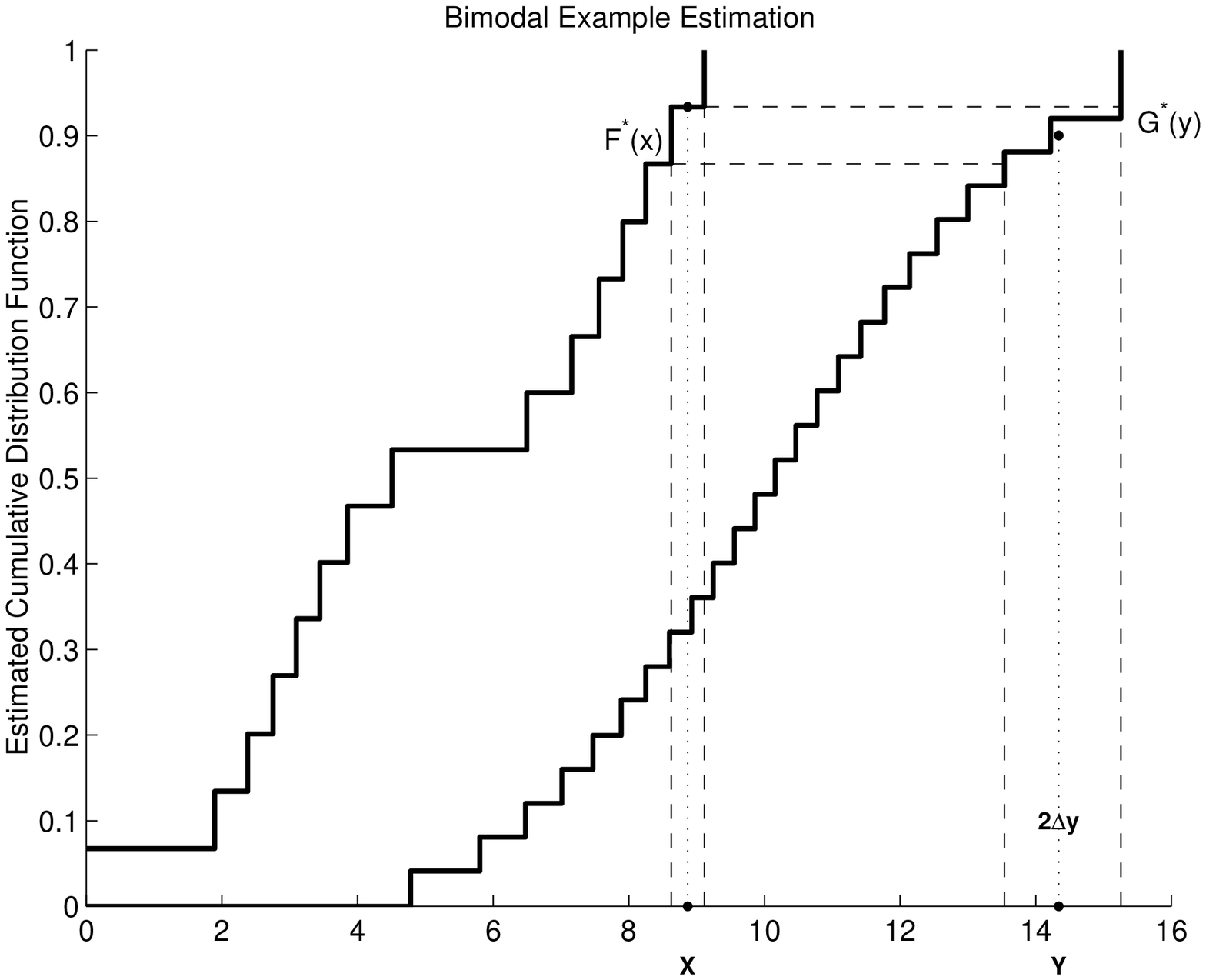}}
}

\caption{The constructed $F^{\ast}(x)$ and $G^{\ast}(y)$ are plotted on the
same
horizontal axis. We have assumed only \Nx\  samples from $X$ and \Ny\ samples
from $Y$. The width $2\Delta y$ represents the uncertainty in the estimates of
$Y = u(X)$. The estimation error grows in the tails of the distributions, owing
to under-sampling of the low probability density.}
\label{UncertaintyFig}
\end{figure}

Henceforth, we will only treat the case $u'(X) > 0$, but the interested
reader can easily derive the $u'(X) < 0$ case in a similar fashion. 
To obtain $u(X)$ for a particular $X$, we find $i$ such that
\begin{equation}
x_{i} \leq X < x_{i+1}. \label{Iconditions}
\end{equation}
\noindent
 Next we find $j_{1}$ and $j_{2}$ such that
\begin{eqnarray}
G^{\ast}(y_{j_{1}}) &\leq& F^{\ast}(x_{i}) < G^{\ast}(y_{j_{1}+1}),
\label{J1condition} \\*
G^{\ast}(y_{j_{2}}) &\geq& F^{\ast}(x_{i+1}) > G^{\ast}(y_{j_{2}-1}).
\label{J2condition}
\end{eqnarray}
\noindent
We then have an estimate of $Y$
\begin{equation}
Y \approx \frac{y_{j_{1}}+y_{j_{2}}}{2}.
\label{Yestimator}
\end{equation}
\noindent
By determining a $Y$ for each sample of $X$, we achieve a tabular
definition of $Y = u(X)$.
We have depicted the mapping process and the uncertainty for the bimodal
example
in \callout{Figure~\ref{UncertaintyFig}}. We have chosen artificially small
datasets of $N_{x}=\Nx$ and $N_{y}=\Ny$ to illustrate the estimation
effect. 

The
approximate uncertainty $\Delta y$ in the $Y$ estimated 
from (\ref{Yestimator}) is given by
\begin{equation}
\Delta y \approx \frac{y_{j_{2}}-y_{j_{1}}}{2}.
\label{rawDeltaYeqn}
\end{equation}
\noindent
We can rewrite (\ref{rawDeltaYeqn}) in terms of $G^{\ast-1}$ as
\begin{eqnarray}
\Delta y &\approx&
\frac{G^{\ast-1}(j_{2}/N_{y})-G^{\ast-1}(j_{1}/N_{y})}{2} \nonumber \\*
&=&
\frac{j_{2}-j_{1}}{2N_{y}}\frac{G^{\ast-1}(j_{2}/N_{y})-G^{\ast-1}(j_{1}/N_{y})}
{(j_{2}-j_{1})/N_{y}}.
\label{DeltaInG}
\end{eqnarray}
This expression contains a first order estimate of the derivative of
$G^{\ast-1}$ which, using (\ref{yCDF}), can be expressed in terms of $g(y)$ as
\begin{eqnarray}
\frac{G^{\ast-1}(j_{2}/N_{y})-G^{\ast-1}(j_{1}/N_{y})}{(j_{2}-j_{1})/N_{y}}
\approx D(G^{\ast-1}) \nonumber \\*
 \approx D(G^{-1}) = \left(\frac{dG}{dy}\right)^{-1} =
\frac{1}{g(y)}.
\end{eqnarray}
Therefore (\ref{DeltaInG}) can be expressed as
\begin{equation}
\Delta y \approx \frac{j_{2}-j_{1}}{2N_{y}g(\frac{y_{j_{1}}+y_{j_{2}}}{2})}.
\end{equation}
Rewriting (\ref{J1condition}) and (\ref{J2condition}) using (\ref{Fest})
and (\ref{Gest}), we have
\begin{eqnarray}
\frac{j_{1}}{N_{y}} &\leq& \frac{i}{N_{x}}, \\*
\frac{j_{2}}{N_{y}} &\geq& \frac{i+1}{N_{x}}.
\end{eqnarray}
\noindent
Therefore
\begin{equation}
j_{2}-j_{1} \geq \frac{N_{y}}{N_{x}},
\end{equation}
\noindent
which leads us to
\begin{equation}
\Delta y \stackrel{>}{\sim}
\frac{1}{2g(\frac{y_{j_{1}}+y_{j_{2}}}{2})N_{x}} = \frac{1}{2g(y)N_{x}}.
\label{DeltaYeqn}
\end{equation}
\noindent
Here, $N_{x}$ accounts for the sampling effect. This relationship implies
that in
the rarified regions of the $Y$ distribution, where $g(y)$ is small, the
estimation error is large. It also suggests that, to first order,
increasing the
$Y$ sample size $N_{y}$ is not as useful in reducing $\Delta y$ as would be
increasing the $X$ sample size $N_{x}$. However, the uncertainty in $x$ is also
important because the total uncertainty in $x$-$y$ space is $\Delta x
\Delta y$.
By a derivation similar to that of $\Delta y$, we have
\begin{equation}
\Delta x \stackrel{>}{\sim} \frac{1}{2f(x)N_{y}}, \label{DeltaXeqn}
\end{equation}
for a total uncertainty of
\begin{equation}
(2\Delta x) (2\Delta y) \stackrel{>}{\sim} \frac{1}{f(x)g(y)N_{x}N_{y}}.
\end{equation}
\noindent
To improve the overall quality of the reconstruction of $Y = u(X)$, we
would like
both $N_{x}$ and $N_{y}$ to be as large as possible.

\section{The SAR on Noisy Data}
Another consideration for the implementation of the SAR method is the effect of
noise. Until now, we have assumed that there is no noise in our measurements
of $X$ and $Y$. However, in practice, we always encounter noisy data, and we
want to be sure that the SAR does not become invalid under typically 
noisy conditions.
In a standard regression, where simultaneous values of $X$ and $Y$ are
known, a least-squares approach can be used to determine $u(X)$ from noisy $X$
and $Y$. We will attempt to demonstrate the effect of noise on the SAR method
by simulating a noisy version of the meteorological example. We generate 100 
noisy samples from the
distributions $f(x)$ and $g(y)$ given in (\ref{AnalyticalPDFx}) and
(\ref{AnalyticalPDFy}). The noise distributions are chosen to be unbiased
Gaussians with standard deviations $\eta_{x}$ and $\eta_{y}$ for $X$ and $Y$.
For now, we choose $\eta_{x}$ and $\eta_{y}$ to be 25\% of the standard
deviations $\sigma_{x}$ and $\sigma_{y}$ of $X$ and $Y$. We can fit the noisy
data with $\log Y = \alpha X+\log \beta$. We perform two such fits: a standard
least-squares regression on the $(X,\log Y)$ pairs and least-squares regression on 
the $(X,\log u(X))$ pairs produced by
the SAR method described in (\ref{Yestimator}). For this parametric example, 
a maximum likelihood estimation of $\alpha$ and $\beta$ would probably
outperform the least-squares approach, but we will compare to the more familiar
regression for this illustration.

\begin{figure}
\showfigs{
\newpage
\centerline{\epsffile{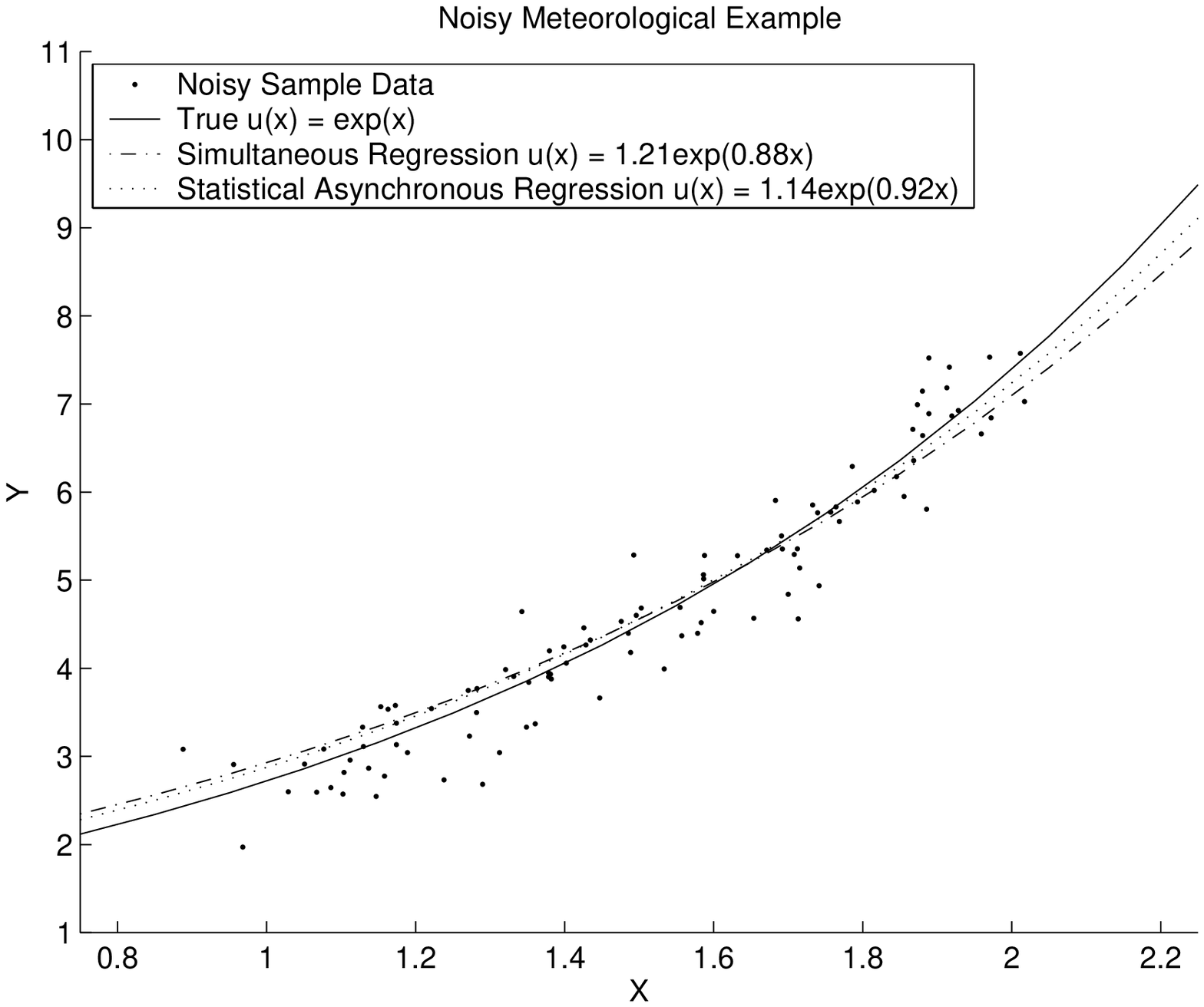}}
}

\caption{Two approximations to the true $u(X)$ show the effect of noisy
samples.
In this simulation, the SAR approximation is actually closer to the true
$u(X)$
than a simultaneous regression.}
\label{PedanticNOISE1}
\end{figure}

Ideally, $\alpha=1$ and $\beta=1$, but, for the noisy data, 
the two regressions give
\begin{eqnarray}
u(x) = 1.21e^{0.88x} & {\rm Standard~Regression}, \\*
u(x) = 1.14e^{0.92x} & {\rm SAR}.
\end{eqnarray}
\noindent
The SAR fit is significantly better than the standard regression. In the
future, it would be
interesting to study how this depends on the type of noise and the form of
$u(X)$. In
\callout{Figure~\ref{PedanticNOISE1}}, we see a graphical depiction of the noisy
data and the two fits. Both fits lie very close to the 
true $u(X)$ curve compared to
the noisy data, however there is a clear improvement with the SAR fit.

\begin{figure}
\showfigs{
\newpage
\centerline{\epsffile{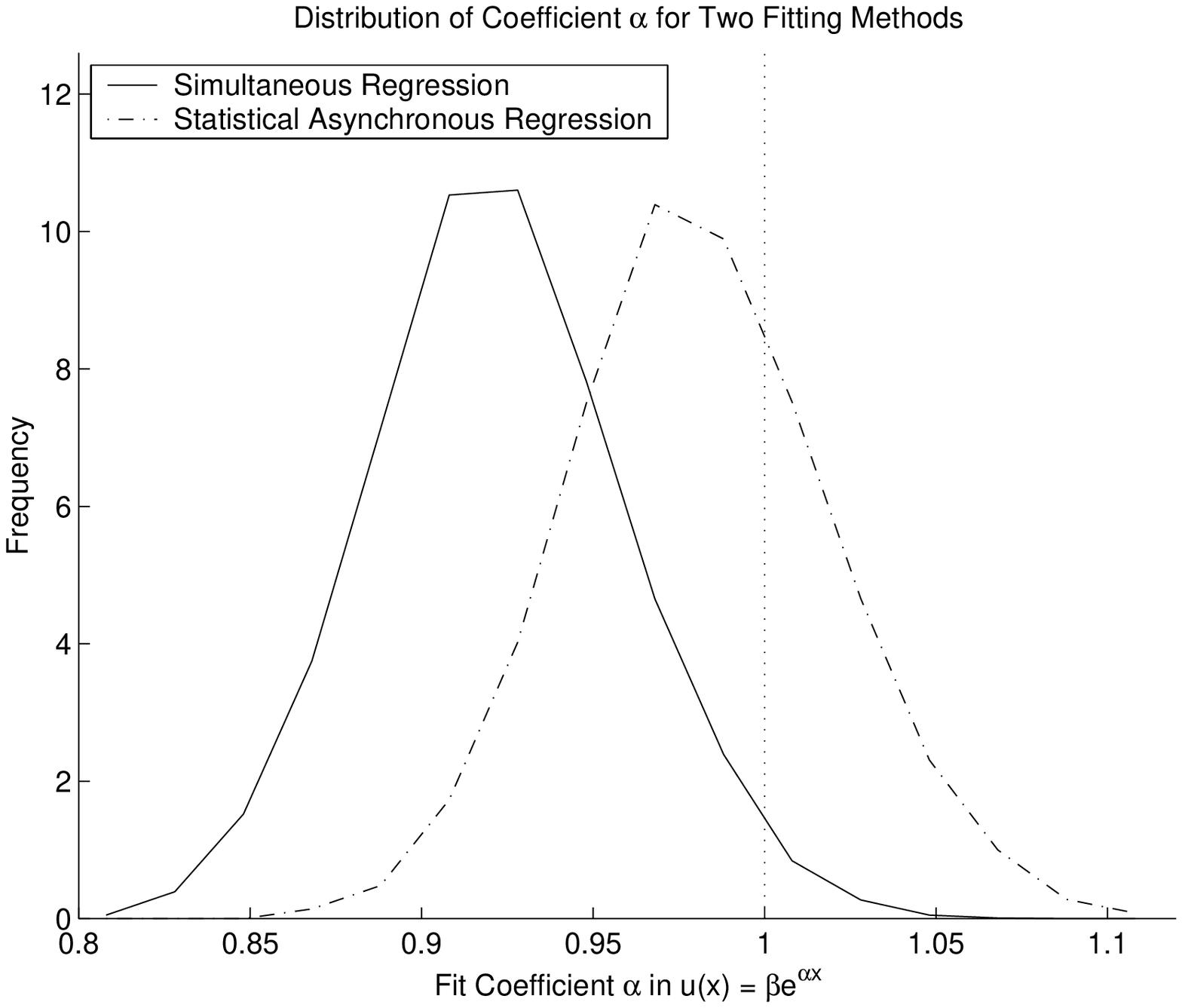}}
}

\caption{The distributions of values of $\alpha$ in
$u(x) = \beta e^{\alpha x}$ obtained
from two regression methods. For this example, the SAR method typically
produces a better
$\alpha$ than a standard simultaneous regression.}
\label{PedanticNOISE2}
\end{figure}

\begin{figure}
\showfigs{
\newpage
\centerline{\epsffile{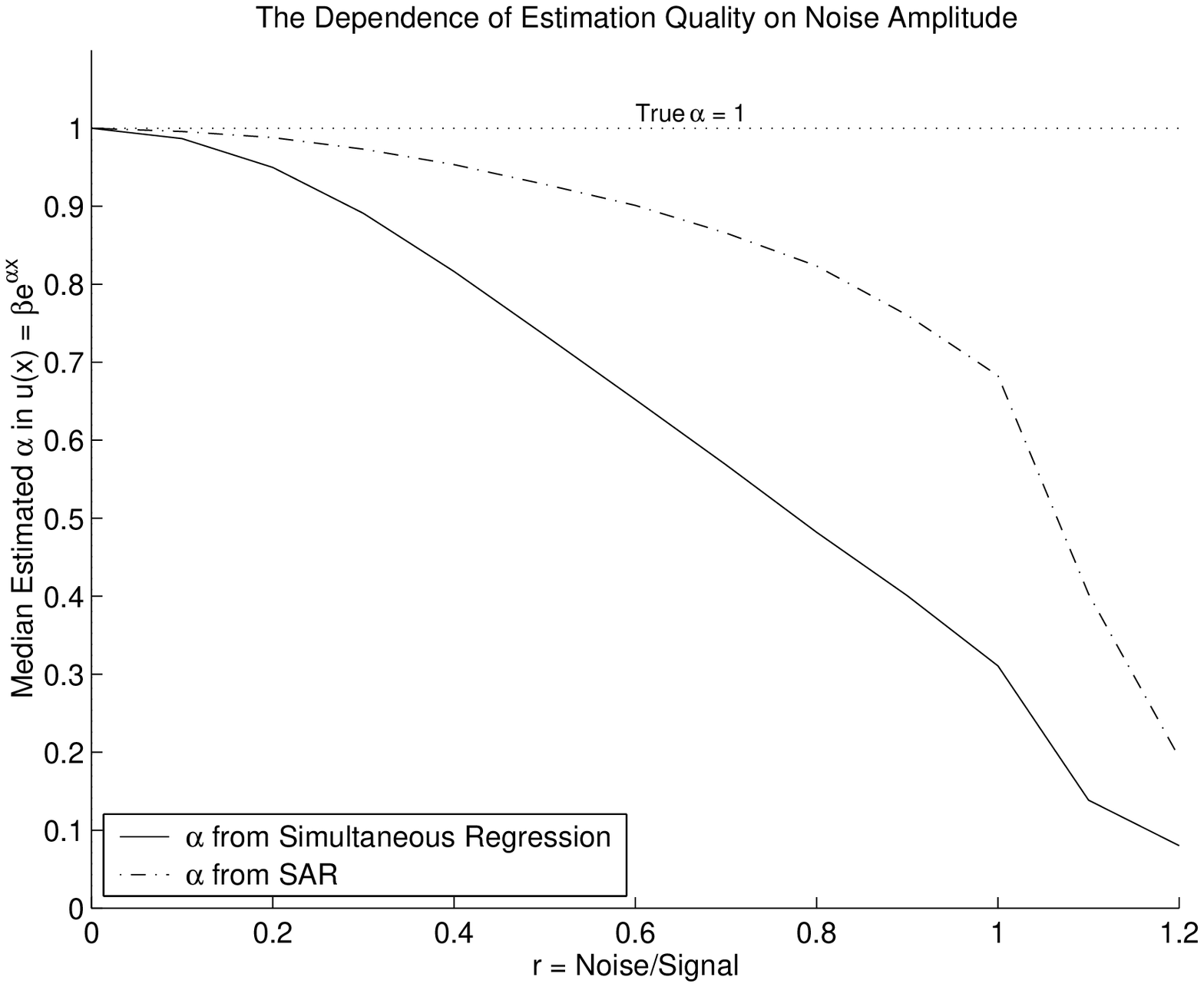}}
}

\caption{The median estimated $\alpha$ in $u(x) = \beta e^{\alpha x}$ for
two regression
methods. The SAR estimate quality drops more slowly than that of the
simultaneous regression.}
\label{PedanticNOISE3}
\end{figure}

To understand better the effect of noise, we repeat the above simulation 5000
times to obtain a distribution of $\alpha$ for each fitting approach. These
distributions are plotted in \callout{Figure~\ref{PedanticNOISE2}}. It is clear
that both methods provide biased estimates of $\alpha$. The SAR method 
produces a
smaller bias, but we would still like to know how that bias depends on the
noise
amplitude. We can test this dependence by finding the bias for various
noise/signal ratios $r$. We will choose the same $r$ for $X$ and $Y$, such that
\begin{equation}
r = \frac{\eta_{x}}{\sigma_{x}} = \frac{\eta_{y}}{\sigma_{y}}.
\end{equation}
So far, we have only tested $r=0.25$, but now we will test a full range from
$r=0$ to $r=1.2$. In \callout{Figure~\ref{PedanticNOISE3}}, we have plotted the
median estimated $\alpha$ versus $r$. We see that for small noise, the estimate
quality is high, but, as $r$ approaches 1, the estimation fails. The $\alpha$
estimated by the SAR method is generally of higher quality than the estimate 
from the
standard regression. For relatively large noise amplitude neither regression
method produces quality estimates of $\alpha$. It is clear that, while the
derivation of the SAR method assumes noiseless data, our implementation
of the SAR method is at least as robust to noise as is the traditional 
least-squares regression. The SAR appears to be reliable when the noise
amplitude is small compared to the variability of the data sample.

\section{An Example from Space Physics}

Finally, we would like to demonstrate the SAR method on a real problem from space
physics. The GOES~8 geosynchronous spacecraft measures, among other things, the
flux of electrons with energies above 2 MeV. The spacecraft orbits the
Earth once per day. The electron populations at geosynchronous orbit are
organized by the position of the Sun relative to the Earth, which we
identify as
local time. Owing to the asymmetry of the Earth's magnetic field in space,
as the
spacecraft passes through different local times, it measures slightly different
parts of the radiation belts. Because the relativistic electron density varies
smoothly with altitude and the electrons themselves make slightly
elliptical orbits every few minutes, hour averaged fluxes at all locations around
geosynchronous orbit are well correlated \cite{li97}; we therefore expect 
a monotonically increasing function $Y = u(X)$ relating fluxes $X$ measured
at one local
time $lt_{x}$ to fluxes $Y$ measured at another local time $lt_{y}$. We can
estimate the flux at $lt_{y}$ from a measurement made by the spacecraft at
$lt_{x}$ if we can determine $u(X)$. The probability distributions of electron
measurements at every local time at geosynchronous are relatively stationary in
time; that is, the distribution of measurements in one year is roughly
equivalent
to the distribution of measurements in any other year. Therefore, we can
estimate
$F(x)$ and $G(y)$ using historical measurements of $X$ and $Y$, and we
can use the
SAR method to reconstruct $u(X)$. We will assume $lt_{x}$ is local 
dawn and $lt_{y}$ is local noon.

We have obtained GOES~8 measurements for 1998 from CDAWeb
(http://cdaweb.gsfc.nasa.gov/) \cite{mcguire2000}. 
We calculated hourly averages and grouped them
into 1-hour bins near local dawn and local noon. This gives us about 360
samples
at each location, but none that are simultaneous because the spacecraft
is only at one location at a time. Because electron measurements tend to 
be heavily biased toward
low values, we will use the Complementary Cumulative Distribution Functions
$F_{>}(x) = 1-F(x)$ and $G_{>}(y) = 1-G(y)$. In terms of these functions, for a
monotonically increasing $u(X)$, we have
\begin{equation}
u(x) = G^{-1}(F(x)) = G_{>}^{-1}(F_{>}(x)). \label{UforUposComplements}
\end{equation}

\begin{figure}
\showfigs{
\newpage
\centerline{\epsffile{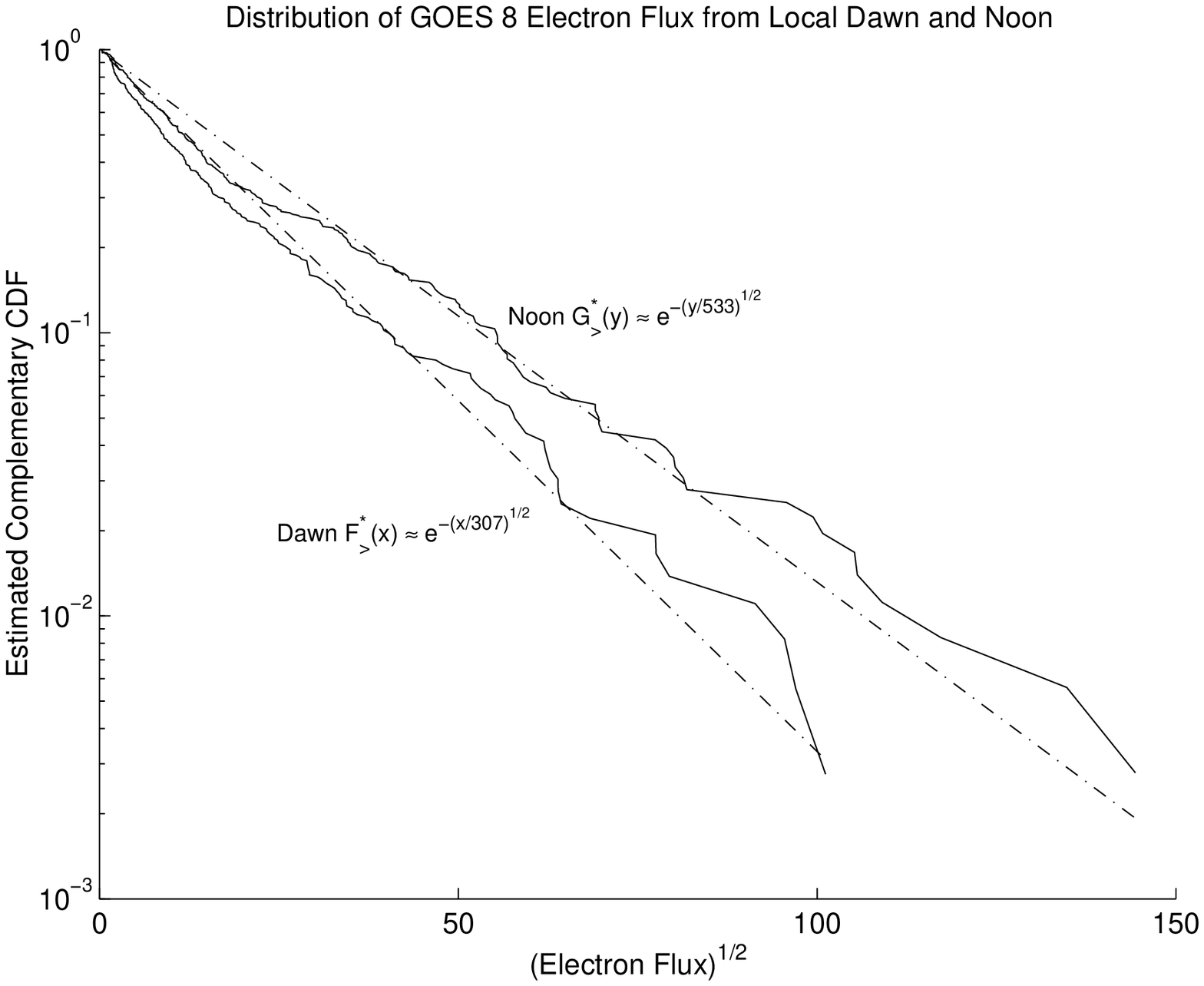}}
}

\caption{This plot depicts estimated complementary CDFs for $X$ and $Y$.
Note that the
vertical axis is logarithmic and the horizontal axis is to the \case{1}{2}
power. The
solid lines represent the tabular forms of the CDFs and the dashed lines
depict the analytical fits with equations (\ref{tre}) and (\ref{hfs}).}
\label{ElectronCDF}
\end{figure}

\begin{figure}
\showfigs{
\newpage
\centerline{\epsffile{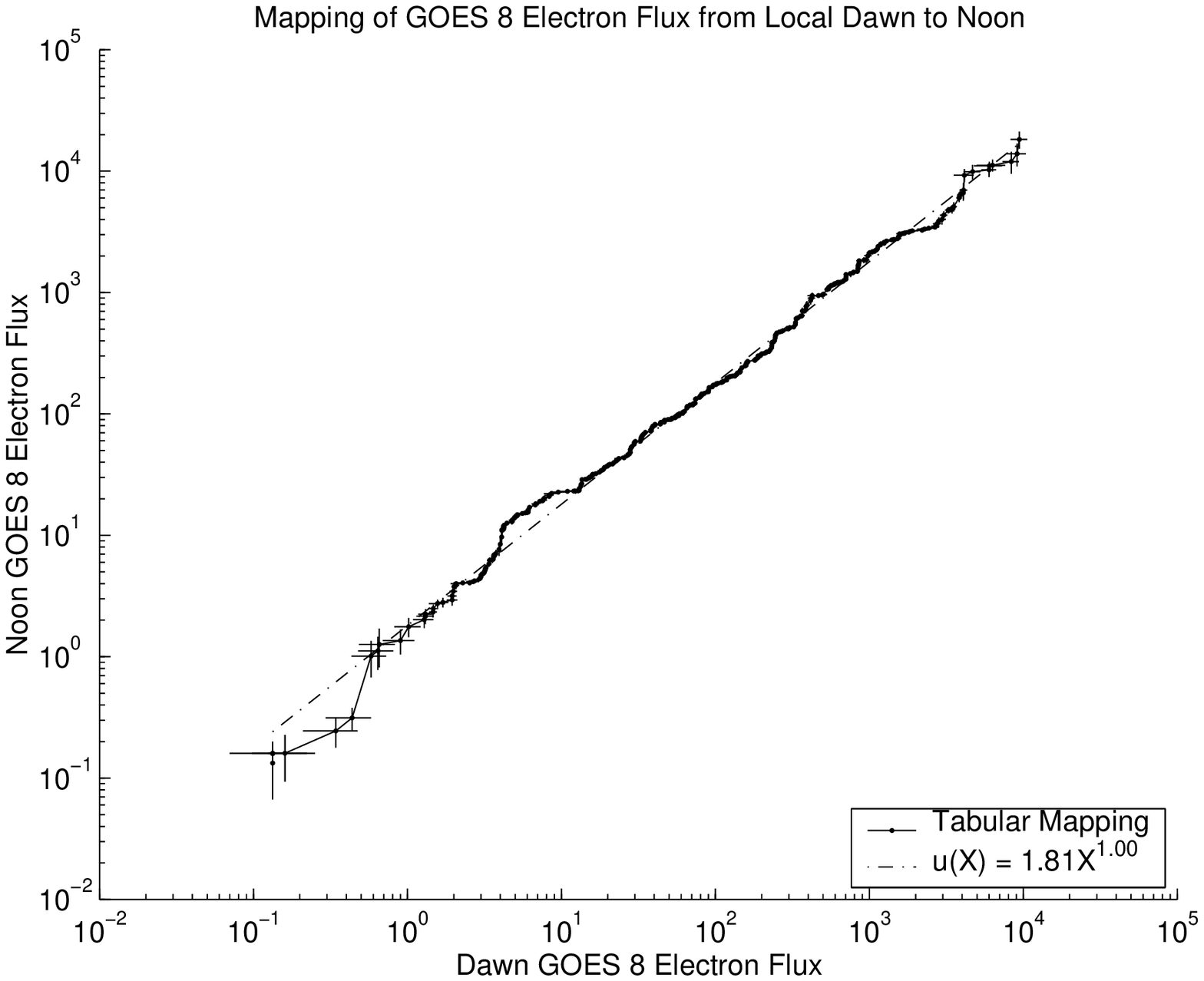}}
}

\caption{The figure shows the tabular and analytical representations of the
mapping function
from dawn flux to noon flux. It is impossible to obtain this function using
simultaneous
measurements because GOES~8 is never both at dawn and at noon. The crosses
through
each dot have been exaggerated to indicate $\pm 3 \Delta x$ and $\pm 3
\Delta y$ as calculated by (\ref{DeltaXeqn}) and (\ref{DeltaYeqn}). The plot
indicates a simple
proportional mapping from $X$ to $Y$, which is physically very reasonable.}
\label{ElectronMapping}
\end{figure}

\callout{Figure~\ref{ElectronCDF}} shows the constructed $F^{\ast}_{>}(x)$ and
$G^{\ast}_{>}(y)$. We can fit both distributions with the same analytical form:
\begin{eqnarray}
F^{\ast}_{>}(x) \approx e^{-\sqrt{\frac{x}{307}}}
~~~~{\rm (Dawn)},  \label{tre} \\*
G^{\ast}_{>}(y) \approx e^{-\sqrt{\frac{y}{533}}}
~~~~{\rm (Noon)}.  \label{hfs}
\end{eqnarray}
\noindent
Assuming an increasing $u(X)$, we use (\ref{UforUposComplements}) to arrive
at an analytical form for $u(X)$:
\begin{eqnarray}
u(X) &=& G^{\ast-1}_{>}\left(F^{\ast}_{>}(x)\right) \nonumber \\*
&=& 533\left(-\log F^{\ast}_{>}(X) \right)^{2} = \frac{533}{307} X \nonumber \\*
&=& 1.74X. \label{nfak}
\end{eqnarray}
The non-parametric SAR mapping is shown in \callout{Figure~\ref{ElectronMapping}} 
to be nearly a power-law. We have determined an analytical fit to be
\begin{equation}
Y = u(X) = (1.8 \pm 0.4)X^{1.00 \pm 0.04}.  \label{fakal}
\end{equation}
\noindent
This fit is in agreement with the function $u(X) = 1.74X$ derived in
(\ref{nfak}) above from
the implementation of the SAR method using the parameterizations
(\ref{tre}) and (\ref{hfs}) of the cumulative distributions. 

The fact
that the exponent in (\ref{fakal}) is very nearly 1 indicates that the densities 
at dawn and noon
change in fixed proportion to each other, even as the radiation belts are
filled during geomagnetic activity. 
If we imagine the electron phase-space structure to
be $f(\vec{r},\vec{v},t)$, then we can state the proportionality as
\begin{equation}
f(\vec{r}_1,\vec{v}_1,t) \propto f(\vec{r}_2,\vec{v}_2,t).
\end{equation}
\noindent
This relationship suggests a simple separation of variables:
\begin{equation}
f(\vec{r},\vec{v},t) = \tilde{f}(\vec{r},\vec{v}) N(t),
\end{equation}
\noindent
where $\tilde{f}(\vec{r},\vec{v})$ represents a phase space shape function, 
and $N(t)$ represents the varying global relativistic electron content of the 
geosynchronous region.

The parameterizations (\ref{tre}) and (\ref{hfs})
together with the corresponding prediction (\ref{nfak}) validated by the direct
non-parametric implementation of the SAR method giving (\ref{fakal}) suggest in
addition
a simple and useful representation of the heavy tail structure of the
distribution of
electron fluxes in terms of stretched exponentials. Such distributions have
been
found to parameterize a large variety of distributions found in
nature as well as in social sciences \cite{Lahsor}. They present a
quasi-stable property \cite{portfolio,book} and can be shown to be the generic
result of the product of random variables in the ``extreme deviation'' regime
\cite{frischsor}.

So far, we have only determined the mapping from local dawn to local
noon.  It
may also be necessary to allow u(X) to vary with magnetic activity level. The
magnetic indices Dst and Kp measure the intensity of the magnetospheric ring 
current and the variability of magnetospheric currents, respectively 
\cite{mayaud80}. We can create different mappings $u(X;Dst,Kp)$ for each of 
several bins of geomagnetic indices; such binning would organize the
data by
the state of the system, reinforcing the assumption that each 
$u(X;Dst,Kp)$ is monotonic and time invariant.
Using the SAR method, we can find mapping functions from every local time to
every other local time, depending on geomagnetic activity, as necessary; this
allows us to reconstruct the flux around the entire orbit at any time based
only
on the single measurement made by GOES~8. If we produce fluxes around the
entire
orbit every hour, we can view  spatial and temporal variations separately. In
particular, if we reconstruct a time series of hourly fluxes at a fixed local
time, we can perform various time series analyses that will not be
influenced by
the spatial variations seen in the measured time-series. This investigation
will be reported elsewhere.

\section{Discussion}

We have shown that it is possible to accurately determine the function 
relating two
variables even when they are not measured simultaneously.
Specifically, we were trying to map 
energetic electron fluxes between different local times at geosynchronous 
orbit. However, we believe that our solution may be useful to other researchers 
whose data are not taken simultaneously. We developed a technique, Statistical 
Asynchronous Regression (SAR), that uses the statistical distributions of 
two variables to determine the unique monotonic function that can map one 
distribution onto the other. Because the SAR technique only works when there 
is a monotonic relationship between the two quantities, it should only be 
applied to quantities that are believed to be highly correlated with each 
other. We caution that the SAR technique will produce a relationship for any 
two quantities, regardless of whether they are actually related. It is 
particularly inappropriate to use the SAR to describe chaotic systems, which
generally arise from non-monotonic behaviors. Also, when the noise amplitude
is a substantial fraction of the data sample variability, we do not expect
the SAR to give reliable results.

To illustrate the SAR technique when the two distributions are known 
analytically, we have provided several examples of common distributions. We 
have shown that the SAR technique can recover the underlying relationship of 
the two quantities even when one distribution diverges or has more than one 
local maximum. We have provided a simple algorithm for implementing the SAR. 
We derived simple expressions for the uncertainty in the estimated relationship 
between the two quantities. To ensure that the technique is robust for noisy 
data, we have simulated two noisy variables with a known relationship and 
determined how well the SAR technique recovers that relationship; the SAR 
performs than a least squared error regression, which 
requires simultaneous measurements of both quantities. While we expect that 
ultimately most scientists will wish to fit $u(X)$ to some parametric form, 
we feel that it is important that the SAR does not require us to assume a 
parametric form a priori.

For those wishing to apply the SAR technique to problems where $u'(X)$ passes 
through zero, we offer the following strategy: if the $u'(X)=0$ occurs at 
known $X$ and $Y$, then the SAR technique is perfectly valid in bins of $X$ 
and $Y$ constrained to be between the zeros of $u'(X)$. In this way, the SAR 
would provide a piecewise form of $u(X)$. 

In closing, we would like to suggest some areas that might benefit from 
the SAR approach. In modeling tectonic deformations, it is useful to quantify 
the balance of deformation accommodated by different faults in a complex network 
\cite{Cowie}. For an individual fault, we often can measure only its length or 
its offset. Relying only on faults with both length and offset known would 
exclude many useful measurements. However, the physics of tectonic deformation 
leads us to expect a monotonic relationship between fault length and offset. 
In this case, the SAR technique would allow us to regress fault length against 
fault offset, using all of the available measurements. Similarly, for individual 
earthquakes we often know only one of seismic moment and energy released 
\cite{energymome}; the SAR technique would allow us to regress all the available 
measurements rather than only those from earthquakes with both moment and energy 
known. We hope that the ideas presented here will assist those who need to
relate non-simultaneous measurements.

\appendix
\section{Appendix: The Change of Variables Theorem}

The SAR method relies heavily on the change of variables theorem
from probability theory. The following derivation will be instructive
to those not familiar with the manipulation of probabilities.

We will use the notational style $P[X \leq x]$ to denote the
probability that any sample from the population of $X$ will be less than or
equal
to some threshold $x$. The formal definitions of the probability density 
functions (PDFs) and cumulative distribution functions (CDFs) for $X$ and
$Y$ are:

\begin{eqnarray}
f(x)dx &=& P[x \leq X < x+dx], \label{xPDF} \\*
g(y)dy &=& P[y \leq Y < y+dy], \label{yPDF} \\*
F(x) &=& P[X \leq x] = \int_{-\infty}^{x} f(x')dx', \label{xCDF} \\*
G(y) &=& P[Y \leq y] = \int_{-\infty}^{y} g(y')dy'. \label{yCDF}
\end{eqnarray}
By definition $f(x)$ and $g(y)$ are non-negative, and $F(x)$ and $G(y)$ reach 1
at $+\infty$. For most purposes, $f(x)$ and $g(y)$ are finite, continuous functions,
and we will operate under that assumption. Accordingly, $F(x)$ and $G(y)$ are 
monotonically increasing, invertible functions.

We assume there is a continuous function $u(X)$ that provides the $Y$ that
corresponds to a given $X$,
\begin{equation}
Y = u(X).
\label{Udef}
\end{equation}
\noindent
We will occasionally replace $Y$ and $X$ with $y$ and $x$, but this should
not worry the reader, as the function $u$ has the same meaning regardless of
its argument. This function must also be monotonic:
\begin{equation}
u'(X) \ne 0~~{\rm for~all}~X.
\label{Umono}
\end{equation}
\noindent
Not only does this imply that $u(X)$ is unique and invertible, but it also
implies that the sign of $u'(X)$ must be either always positive or always
negative. Strictly speaking, $u'(X)$ may vanish at isolated
points, so long as it only touches, but does not traverse, zero.

We can write $u'(x)$ as
\begin{equation}
u'(x) = \lim_{\Delta x \rightarrow 0} \frac{u(x+\Delta x)-u(x)}{\Delta x}. \label{Ulim}
\end{equation}
\noindent
If $u'(x)$ is positive for all values of $x$, then $\Delta x > 0$ implies that $u(x+\Delta x) > u(x)$.
Formally, with $x+\Delta x$ replaced by $X$, we state this as
\begin{equation}
u'(x) > 0 \Longleftrightarrow \{X > x \Leftrightarrow u(X) > u(x)\}. \label{Upos}
\end{equation}
\noindent
By similar reasoning,
\begin{equation}
u'(x) < 0 \Longleftrightarrow \{X > x \Leftrightarrow u(X) < u(x)\}. \label{Uneg}
\end{equation}
\noindent

For the case $u'(x) > 0$, we can therefore replace the inequality
in (\ref{xCDF}) according to (\ref{Upos}) to arrive at
\begin{equation}
F(x) = P[X \leq x] = P[u(X) \leq u(x)].
\end{equation}
\noindent
Using (\ref{yCDF}) and (\ref{Udef}), we have
\begin{equation}
F(x) = P[Y \leq u(x)] = G(u(x)). \label{FeqGu}
\end{equation}
\noindent

For the other case, $u'(X) < 0$, we can use (\ref{Uneg}) similarly to
replace the inequality in (\ref{xCDF}), which gives
\begin{eqnarray}
F(x) &=& P[X \leq x] = P[u(X) \geq u(x)] \nonumber \\*
&=& 1-P[u(X) < u(x)]. \label{UnegReplacement}
\end{eqnarray}
\noindent
For a finite, continuous distribution $f(x)$,
$P[u(X) < u(x)] = P[u(X) \leq u(x)]$.
Therefore, we can apply (\ref{yCDF}) and (\ref{Udef}) to
(\ref{UnegReplacement}) to arrive at
\begin{equation}
F(x) = 1-P[Y \leq u(x)] = 1-G(u(x)). \label{Feq1-Gu}
\end{equation}
\noindent

\begin{figure}
\showfigs{
\newpage
\centerline{\epsffile{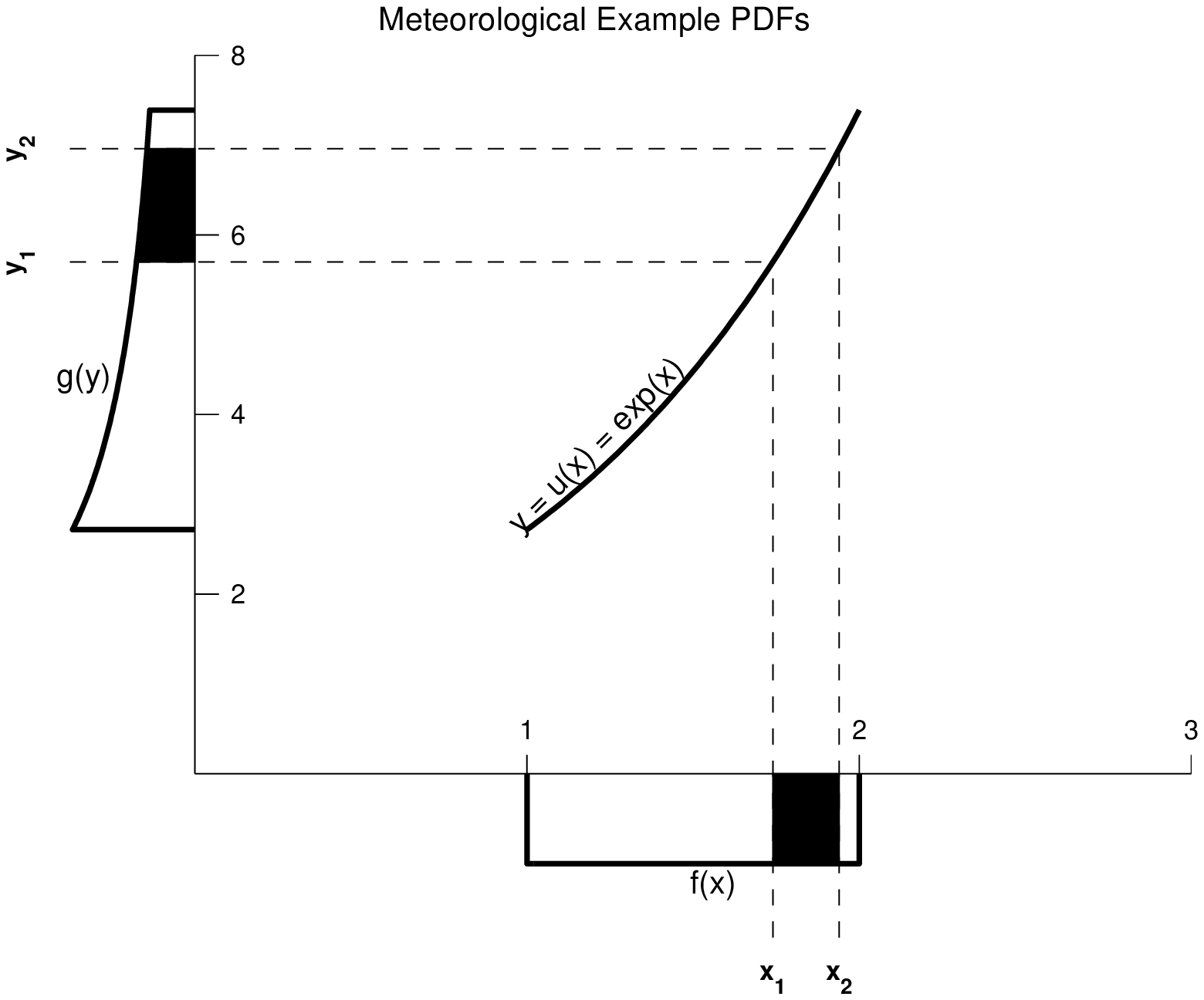}}
}

\caption{Probability densities for $X$ and $Y$ are plotted outside the
respective axes. The relational function $Y = u(X)$
provides the scaling from $X$ to $Y$. Consistent with the conservation of
probability, the shaded regions have equal area.}
\label{PedanticPDF}
\end{figure}

\begin{figure}
\showfigs{
\newpage
\centerline{\epsffile{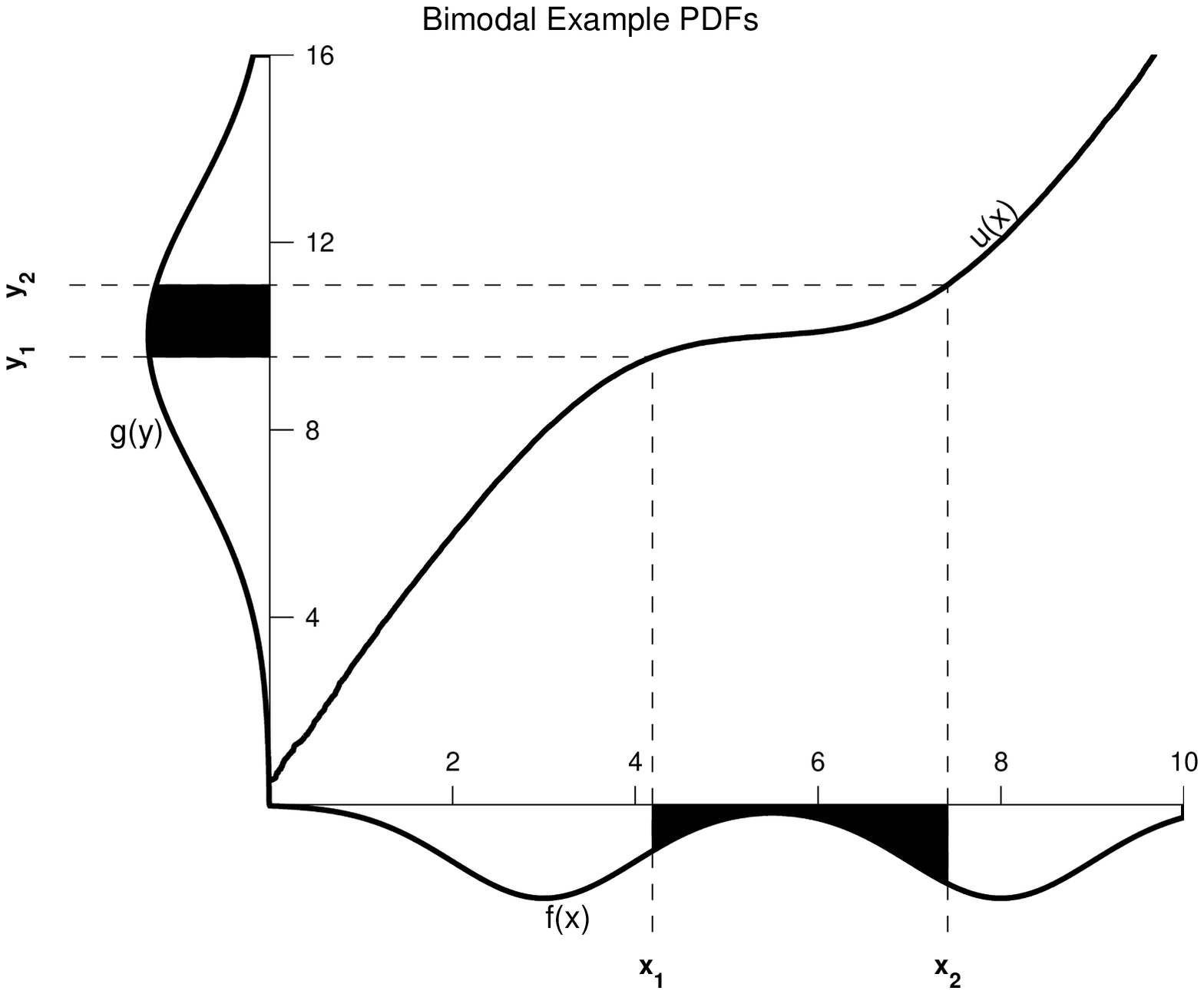}}
}

\caption{In the same format as Figure~\ref{PedanticPDF}, this is a
depiction of the mapping from a bimodal to a Gaussian.
The SAR method easily handles the bimodal $f(x)$ and the highly non-linear
$u(x)$.}
\label{BimodalPDF}
\end{figure}

\begin{figure}
\showfigs{
\newpage
\centerline{\epsffile{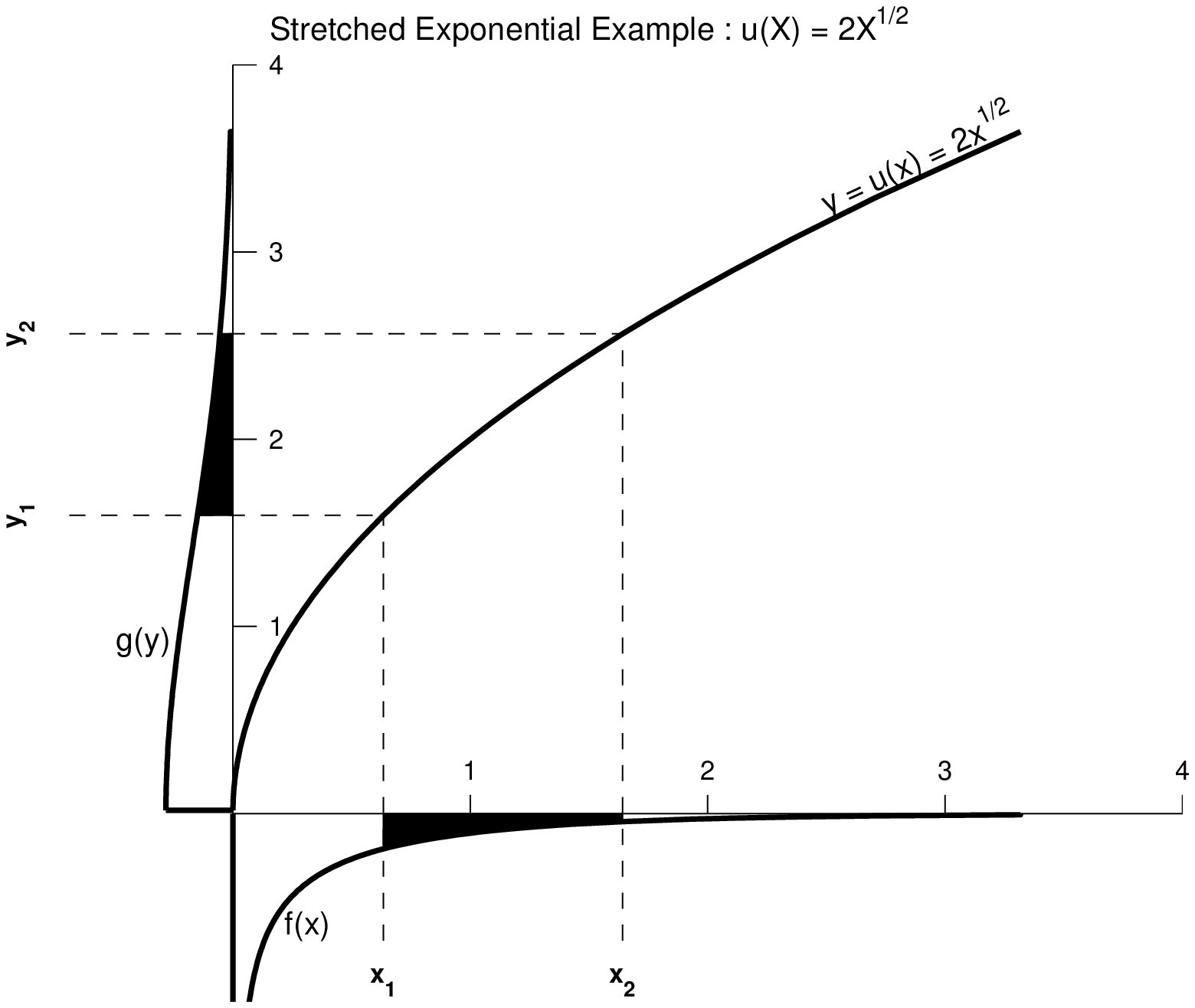}}
}

\caption{In the same format as Figure~\ref{PedanticPDF}, this depicts the
mapping
from a stretched exponential to a Gaussian. The divergence in $f(x)$ does
not prevent the SAR method from recovering  $u(x)$.}
\label{StretchPDF}
\end{figure}

By differentiating (\ref{FeqGu}) and (\ref{Feq1-Gu}), we arrive at
\begin{eqnarray}
f(x)dx = g \left( u(x) \right) u'(x)dx~~{\rm for}~~u'(x) > 0, \\
f(x)dx = -g \left( u(x) \right) u'(x)dx~~{\rm for}~~u'(x) < 0,
\end{eqnarray}
\noindent
or, equivalently,\\
\makebox[\columnwidth]{$f(x)dx = g(u(x))|u'(x)|dx$ }
\mbox{\makebox[\columnwidth]{$~~~~~~~~~= g(y) \left| \frac{dy}{dx} \right| dx = g(y)|dy|$.}
\makebox[0in][r]{\hspace{-0.5in}(\ref{ChangeOfVars})}} \\
Probability is conserved under a change of variables.
This is the change of variables theorem, and it is depicted graphically
in Figure~\ref{PedanticPDF}. 

\acknowledgements
This work was in part funded by IGPP grant LANL~1001 and NSF grant 
\mbox{ATM 98-19935}. We would like to
thank G.~Reeves
and the Energetic Particles group at Los Alamos National Lab for their
insightful
critique of preliminary presentations of the SAR method. We would also
like to thank CDAWeb and T.~Onsager for providing data from the GOES~8
spacecraft. We thank V.~Pisarenko, F.~Schoenberg, and A.~Russell for 
helpful comments on the technique and manuscript. IGPP~No.~5471.

{}

\end{article}

\begin{thebibliography}{}

\bibitem[{\it Allen and Yen,} 1979]{Allen1979} 
Allen, M.J., and W.M. Yen,
{\it Introduction to measurement theory},  
Brooks/Cole Publishing, Monterey, CA, 1979.

\bibitem[{\it Brautigam et al.,} 1992]{CRRESELE}
\reference Brautigam, D.H., M.S. Gussenhoven, and E.G. Mullen, 
Quasi-static Model of Outer Zone Electrons, 
{\it IEEE Trans. Nucl. Sci., 39,} 1797-1803, 1992.

\bibitem[{\it Cowie and Scholz,} 1992]{Cowie}
\reference Cowie, P.A. and C.H. Scholz, Displacement length scaling
relationship for
faults -- Data synthesis and discussion, {\it J. Struct. Geol. 14},
10:1149-1156, 1992.

\bibitem[{\it Fisher,} 1983]{Fisher83} Fisher, N. I.,
Graphical methods in nonparametric statistics: a review
and annotated bibliography,  {\it International Statistical Review 51}, 
25-58, 1983.

\bibitem[{\it Friedel et al.,} 1999]{friedel99}
\reference Friedel, R.H.W., G. Reeves, D. Belian, T. Cayton,
C. Mouikis, A. Korth, B. Blake, J. Fennell, S. Selesnick,
D. Baker, T. Onsager, and S. Kanekal,
A multi-spacecraft synthesis of relativistic electrons in the
inner magnetosphere using LANL, GOES, GPS, SAMPEX, HEO, and POLAR, 
{\it Radiation Measurements 30}, 589-597, 1999.

\bibitem[{\it Frisch and Sornette,} 1997]{frischsor} Frisch, U. and D.
Sornette,
Extreme deviations and applications,  {\it J. Phys. I France 7}, 1155-1171,
1997.

\bibitem[{\it Hardle,} 1990]{Hardle} Hardle, W.,
{\it Applied nonparametric regression, Econometric Society Monogr.},  
Cambridge University Press, New York, 1990.

\bibitem[{\it Karlen,} 1998]{karlen} Karlen, D.,
Using projections and correlations to approximate probability
distributions, {\it Computer in Physics 12}, 380-384, 1998.

\bibitem[{\it Laherr\`ere and Sornette,} 1998]{Lahsor} Laherr\`ere, J. and
D. Sornette,
Stretched exponential distributions in Nature and Economy: ``Fat tails''
with characteristic scales, {\it European Physical Journal B 2}, 525-539, 1998.

\bibitem[{\it Li et al.,} 1997]{li97}
\reference Li, Xinlin, D.N.
Baker, M. Temerin, T.E. Cayton, E.G.D. Reeves, R.A. Christensen,
J.B. Blake, M.D. Looper, R. Nakamura, and S.G. Kanekal,
Multisatellite observations of the outer zone electron variation during the
November 3-4, 1993 magnetic storm, \jgr {\it 102}, 14,123-14,140, 1997.

\bibitem[{\it Mayaud,} 1980]{mayaud80}
\reference Mayaud, P.N., {\it Derivation, meaning and use of geomagnetic
indices,
Geophys. Monogr. Ser.}, vol. 22, American Geophysical Union, Washington,
D.C., 1980.

\bibitem[{\it Mayeda and Walter,} 1996]{energymome}
\reference Mayeda, K. and W.R. Walter, Moment, energy, stress drop and
source spectra of western United
States earthquakes from regional coda envelopes, \jgr {\it 101},
B5:11195-11208, 1996.

\bibitem[{\it Moorer,} 1999]{Moorer1999}
\reference Moorer, D.,
Specifying outer belt electrons by data assimilation, 
Spring AGU Meeting, 1999.

\bibitem[{\it McGuire et al.,} 2000]{mcguire2000}
\reference McGuire, R.E., R.J. Burley, R.M. Candey, R.L. Kessel,
T.J. Kovalick,
CDAWeb and SSCWeb: Enabling correlative international
sun-earth-connections science entering the era of IMAGE and Cluster, 
Spring AGU Meeting, 2000.

\bibitem[{\it Moran,} 1969]{Moran} Moran, P.A.P.,
 Statistical inference with bivariate gamma distributions,
{\it Biometrika 56}, 627-634, 1969.

\bibitem[{\it Press et al.,} 1992]{Press92} 
Press, W.H., S.A. Teukolsky, w.T. Betterling, B.P. Flannery,
{\it Numerical Recipes in C: The Art of Scientific Computing}, 
Cambridge University Press, Cambridge, MA, 1992.

\bibitem[{\it Reeves et al.}, 1998]{Reeves1998}
\reference Reeves, G. D., D. N. Baker, R. D. Belian, 
J. B. Blake, T. E. Cayton, J. F. Fennell, R. H. W. Friedel, 
M. M. Meier, R. S. Selesnick, and H. E. Spence, 
The global response of relativistic radiation belt electrons to the 
January 1997 magnetic cloud, 
{\it Geophys. Res. Lett., 25}, 3265-3268, 1998.

\bibitem[{\it Selesnick and Blake}, 2000]{Selesnick2000}
\reference Selesnick, R.S. and J.B. Blake, 
On the source location of radiation belt relativistic electrons, 
{\it \jgr, 105}, 2607-2624, 2000.

\bibitem[{\it Sornette,} 2000]{book} Sornette, D., {\it Critical Phenomena
in Complex Systems,
Concepts and Tools for Variability at Many Scales}
(Springer, Heidelberg, 2000)

\bibitem[{\it Sornette et al.,} 2000]{portfolio}
Sornette, D.,  P. Simonetti, and J.V. Andersen,
$\phi^q$-field theory for Portfolio optimization: ``fat tails'' and
non-linear correlations, {\it Physics Reports, 335},
19-92, 2000.


\bibitem[{\it Vette,} 1991]{ae8min}
\reference Vette, J., 
The AE-8 trapped electron model environment, 
National Space Science Data Center, Report 91-24, 
Greenbelt, Maryland, 1991.

\bibitem[{\it Wilk and Gna\-nade\-si\-kan,} 1968]{Wilk68} Wilk, M.B. and
R. Gnanadesikan,
Probability plotting methods for the analysis of data,
{\it Biometrika 55}, 1-17, 1968.


\end{thebibliography}
\end{document}